\documentclass{article}
\usepackage{amssymb}
\usepackage{listings}
\usepackage{booktabs}
\usepackage{algorithm}
\usepackage{algorithmic}
\usepackage{CJK}
\usepackage{tikz}
\usepackage{hyperref}
\usepackage{url}
\usepackage[top=1in, bottom=1in, left=1.25in, right=1.25in]{geometry}
\newcommand{\code}[1]{{\hbox{\small\tt{#1}}}}

\begin{document}
\title{Invariant Detection with Program Verification Tools\footnote{All discussions about tools in this paper are based on our experiments performed during August 2014---November 2014.}  }

\author{Wei He\footnote{This research was conducted under the supervision of Prof. Saumya Debray}
\vspace{1.6mm}\\
\fontsize{10}{10}
Department of Computer Science\\
University of Arizona\\
AZ 85721, USA \\
\fontsize{9}{9}\selectfont\ttfamily\upshape
 hewei@email.arizona.edu
}
\date{}
\maketitle

\begin{abstract}
Compilers can specialize programs having invariants for performance improvement. Detecting program invariants that span large and complex code, however, is difficult for compilers. Traditional compilers do not perform very expensive analysis and thus only identify limited invariants, which limits the potential of subsequent optimizations. We would like to address the invariant detection problem via more sophisticated analyses using program verification tools. In this paper, we reveal pitfalls of choosing program verification tools for invariant detection, identify challenges of modeling program behavior using one of these tools---CVC4, and propose some ideas about how to address the challenges.
\end{abstract}

\section{Introduction}
In this paper, we attempt to address the problem of detecting \textit{invariants}---program properties, such as variable values, that remain unchanged along all possible execution paths between two points in a program. It has been shown that program optimization based on invariants identified via domain knowledge can achieve significant performance improvement~\cite{Zhang3}. However, detecting invariants that span large and complex code is difficult for compilers for two reasons: (1) Compilers usually do not perform expensive analysis due to its performance concern, (2) and therefore can not find all possible invariants in a program. In our research, we would like to explore whether more sophisticated analyses by program verification tools can help us find more invariants that enable further optimization than traditional compilers.  

Compiler optimization has been a effective approach to speedup softwares for a long time. Some of these optimization techniques utilizes invariants in a program source code. In general, invariants are properties that something is known to remain unchanged along all possible execution path of a program. For example, loop-invariant code motion is a compiler optimization that moves computations whose result remain unchanged in the loop body to the outside. The growing size and complexity of programs nowadays, however, makes it difficult for compilers to detect invariants they can utilize. For instance, it is challenging to detect invariants when modern programming language allows a memory location to be accessed through different alias in the program (e.g., pointers in C/C++). Meanwhile, the size of Linux kernal grows from less than 0.5M SLOC to over 2M SLOC within five years from 1994 to 1999~\cite{tu2000evolution} and is over 20M SLOC now.

Since examining properties of programs has been studied in the program verification community as model-checking problems and many tools are available, we would like to detect program invariants by utilizing program verification tools. SAT (Boolean Satisfiability) solver and SMT (Satisfiability Modulo Theories) solver are widely used tools on the model checking problem \cite{bjorner2014applications, sheeran2000checking, vanegue2012smt}. Users need to model their programs using the input language of a solver and give a query about the property they would like to check. The solver then answers the user's queries. Both SMT and SAT solvers answer a query by either proving the queries property is true or finding a counterexample. For SAT solvers, the input needs to be boolean formula in CNF (Conjunctive Normal Form). On the other hand, SMT solvers provide more expressive input languages and is thus easier to use. Considering that real-world softwares can be huge, our first attempt is to model programs using SMT solver for simplicity.

In the paper, we attempt to explore the approach of leveraging sophisticated analysis via an SMT solver to detect invariants in C programs. This paper makes the following contributions:
\begin{itemize}
\item We describe an idea of modeling program behavior using an SMT solver---CVC4 and identify C program structures that are easy or hard to model.
\item We identify pitfalls of choosing program verification tools to detect invariants for optimizations. One needs to pay attention on the approximation taken by the tools because the approximation may affect the correctness of later optimizations.
\item Given that it is hard to model some C structures accurately, we give an initial idea about how to identify the rest of the program not influenced by these structures. Modeling the rest of the program could enable us to detect a subset of invariants in the program.
\end{itemize}

In the rest of this report, we introduce several tools we have tried for invariant detection and the features of each tool in Section~\ref{sec: tools}. Section~\ref{sec: classification} presents how some C language structures can be easily modeled in the input language of an SMT solver---CVC4. In this section, we also discuss structures that are hard to modeled and the specific challenges. Finally, we give an initial idea about how to model a C program with complex structures via a compromise of precision in Section~\ref{sec: heuristic}.

\section{Tools}\label{sec: tools}
For our purpose of detecting invariants in C programs, we consider using SMT solvers, SAT solver, or C program analysis tools built on top of these solvers. As we will discuss in this section, we have two possible options: (1) using a C program analysis tool that takes \textit{over-approximation} on examined properties or (2) modeling the C program behavior in the language of an SMT or SAT solver that proves invariant properties. We explain the requirement of over-approximation and summarize four tools we have tried in this section. 

Some C program analysis tools do not examine all possible execution paths of programs (especially for those with loops) but take an approximation. According to how they take the approximation, these tools fall into two categories: over-approximation and under-approximation. The difference between them is that over-approximation guarantees if the tool gives a positive answer to a query, the true answer of this query is positive indeed. In contrast, under-approximation ensures that if the tool gives a negative answer, there exists a concrete counterexample to the query. For example, suppose there is an integer variable \texttt{int x} in a tool's input and the tool cannot find out a concrete value of \texttt{x}. The user would like to know whether \texttt{x == 0} is true. If the tool takes an over-approximation, it answers \texttt{False} because the value of \texttt{x} is uncertain. If the tool takes an under-approximation, it answers \texttt{True} because the solver cannot find a concrete counterexample. Note that for the reverse question, i.e., whether \texttt{x != 0}, the tool will give the same answer as to the question \texttt{x == 0} because it has no knowledge about the value of \texttt{x}.

For the purpose of invariant detection, we need the tool to take over-approximation because detected invariants are used by subsequent optimizations that must guarantee the correctness. Suppose we would like to determine whether the value of a variable $x$ remains the same after some statements. Let \texttt{old\_x} and \texttt{new\_x} be the value of \texttt{x} before and after the statements, our query to the tool simply ask whether \texttt{new\_x == old\_x} is true or not. If the tool answers true, an optimization may be performed based on this property, so we would like the tool to answer true only if it is provable.

CVC4 \cite{barrett2011cvc4} is an open-source SMT solver that can be either used alone or utilized as a backend of other program analysis tools. CVC4 provides a number of commonly used built-in theories including integer linear arithmetic, arrays, tuples, records, inductive data types, and bit-vectors. When used alone, it takes code in CVC4's native language (we'll call it CVC4-language for brevity) or SMT-LIB as input, and checks the properties specified by queries in the code. CVC4-language does not allow an instance to change its value, which makes it inherent difficult to model loop behavior precisely in CVC4-language. We'll discuss this limitation and how loop behavior can be modeled through induction in Section~\ref{sec:loop2}.

Based on CVC4 and another SMT solver---Z3 \cite{de2008z3}, Cascade \cite{wang2014cascade} is a C program static analysis tool that utilizes these SMT solvers as its backend. It takes as input a C program and a control file that specifies assertions and constraints, and checks the assertions with the backend tools. Cascade takes under-approximation. It unwinds loops and inlines functions in the input program, and reports the assertions to be \textit{safe} if no counterexample can be reached within a fixed number of iterations \cite{cascade@TACAS15}.

Another program verification tool is CBMC \cite{clarke2004tool, clarke2003behavioral}, which implements Bounded Model Checking (BMC) \cite{biere1999symbolic} for ANSI-C/C++ programs. CBMC is based on SAT solver and uses MiniSat2 as the backend in default. It takes a C program as input, and outputs the verification result. Same as Cascade, CBMC takes under-approximation and models functions and loops in the similar way.

Included in the CBMC, a $k$-induction tool provides more powerful verification to programs with loops \cite{donaldson2011software}. It splits a program into two pieces, a base case and a step case, and checks for loop invariants inductively: it checks whether the first $k$ iterations ensure the property and whether the next iteration after $k$ consecutive iterations ensures the property separately.

Unlike CBMC and Cascade which do under-approximating analysis, 2LS is an over-approximation tool \cite{Martin2LS} which takes C programs as input and reports the checking results. 2LS combines together three existed techniques: bounded model checking, $k$-induction, and abstract interpretation, obtaining a new algorithm which is better than any single one of them. Unfortunately, 2LS is still under development and is not robust enough to be applied on very large programs such as {\sc PostgreSQL}.

For invariant detection, all the tools we've tried and described above cannot deal with real-world programs because they take under-approximation or lack robustness. So the rest of this paper describes our effort on modeling C programs in CVC4-language.

\section{Modeling C Structures in CVC4-language}\label{sec: classification}
CVC4-language is very different from C programming languages in that it does not allow an instance to change its value, which makes it difficult to model some C program structures in CVC4-language. In this section, we identify C structures that can be easily modeled in CVC4-language and structures that are difficult to model. We will discuss how to address some difficult C structures in the next section.

CVC4-language provides a number of built-in types, such as integer, real, boolean, array, tuple, record, bit-vector, and function. This makes it easier to model basic type instance and some structures in C program. There are two big differences between CVC4-language and most other programming languages. Firstly, instances in CVC4-language can never change their value after they are defined. Secondly, if an instance is not assigned an initial value when it is defined, it has a non-deterministic value rather than a default value as in C language. To avoid ambiguity, we avoid using ``variable" to refer to an instance in CVC4-language program.

\subsection{Simple structures}

\subsubsection{Data types}
CVC4-language has several built-in types corresponding to the basic types in C program. Specifically, integers, floating point numbers, and boolean variables can be modeled with built-in types \code{INT}, \code{REAL}, and \code{BOOLEAN}. User defined basic type is also supported but as uninterpreted type whose domain is unknown\cite{cvc4language}.

Here is an example of how \code{struct} in C program can be modeled with \code{RECORD} in CVC4-language.

\vspace{2mm}
\code{struct Pair}
\code{\{}
\code{\qquad int x;}
\code{\qquad double y;}
\code{\};}
\vspace{2mm}

The C code above can be represented in CVC4-language as follows.

\vspace{2mm}
\code{Pair : TYPE = [\# x : INT, y : REAL \#];}
\vspace{2mm}

For pointers in C language, we model the memory with one array for each data type used in the program. We then model a C pointer as an integer instance associated with an array index as the memory address by an ASSERT statement in CVC4-language.

\vspace{2mm}
\code{int i;}
\code{double d;}
\vspace{2mm}
\code{int *ip = \&i;}
\code{double *dp = \&d;}
\vspace{2mm}

It can be represented in CVC4-language as follows.

\vspace{2mm}
\code{Memory\_int : TYPE = ARRAY INT OF INT;}
\code{Memory\_real : TYPE = ARRAY INT OF REAL;}
\code{MEMORY : TYPE = [\# mem\_int : Memory\_int, mem\_real : Memory\_real \#];}
\code{mem : MEMORY;}
\vspace{2mm}
\code{i : INT;}
\code{r : REAL;}
\code{ip, dp : INT;}
\code{ASSERT(mem.mem\_int[ip] = i);}
\code{ASSERT(mem.mem\_real[dp] = d);}
\vspace{2mm}

\subsubsection{Operation}
Common arithmetic operators for basic data types are supported in CVC4-language. Besides, one can access members in record or tuple with \textbf{.} (dot) operator such as \code{instance.i} for a tuple where \code{i} is an integer constant indicating the serial number of the member being accessed, or \code{instance.member\_name} for a record.

One can also ``update" an array or a record using WITH operator as follows.

\vspace{2mm}
\code{Array\_new : ARRAY T1 OF T2 = Array\_old WITH [i] := newValue;}
\code{Record\_new : RecordType = Record\_old WITH .member\_name := newValue;}
\vspace{2mm}

\subsubsection{Function}
A function in C can be modeled in CVC4-language as an instance of type \code{(T1, T2, ..., Tn) -> T}, where \code{T1, T2, ..., Tn} on the left-hand side are the types of parameters and \code{T} on the right-hand side is the type of return value. We use \code{LAMDA} expression and \code{LET} statement to specify a concrete function of its type. When modeling functions in C program, we use an instance of the memory array as an extra parameter and part of the return value. This enables changes made on instances in the function body to be visible from outside. An example is given below.

\vspace{2mm}
\code{int GetRowLength(TableHeader table\_header, int index)}
\code{\{}
\code{\qquad int row\_length = 0;}
\code{\qquad row\_length += column\_length[index];}
\vspace{2mm}
\code{\qquad return row\_length;}
\code{\}}
\vspace{2mm}

This C function can be represented in CVC4-language as follows.

\vspace{2mm}
\code{GetRowLength : (TableHeader, INT, MEMORY) -> [INT, MEMORY]}
\code{= LAMBDA(table\_header : TableHeader, index : INT, mem : MEMORY) :}
\code{LET}
\code{\qquad row\_length = 0,}
\code{\qquad row\_length\_new = row\_length + column\_length[index],}
\code{IN (row\_length\_new, mem);}

An exception is that CVC4-language doesn't support recursive function. We'll discuss the problem of recursion in Section~\ref{sec:recursion}.

\subsubsection{Branching}
As one of the mostly used constructs in C program, the if-else branching is supported by CVC4-language. It commonly can be simply modeled with an \textit{IF-THEN-ELSE} statement of the following form where \code{bi} are expressions returning boolean value and \code{ti} are statements.

\vspace{2mm}
\code{IF b1 THEN t1 ELSIF b2 THEN t2 ELSIF ... ENDIF}

\subsubsection{Function pointer}
Function pointer is not supported by CVC4-language, but we can model it with IF-THEN-ELSE statements. The method here is to associate an unique integer with each function as their address, and comparing the function pointer with these possible function addresses to decide which function to call.

A simple example is given below.

\vspace{2mm}
\code{int EqualInt4(unsigned long value1, unsigned long value2)}
\code{\{}
\code{\qquad return (int)value1 - (int)value2;}
\code{\}}
\vspace{2mm}
\code{int LessthanInt8(unsigned long value1, unsigned long value2)}
\code{\{}
\code{\qquad if ((long)value1 < (long)value2)}
\code{\qquad \{}
\code{\qquad \qquad return 0;}
\code{\qquad \}}
\code{\qquad return 1;}
\code{\}}
\vspace{2mm}
\code{typedef int (*PredicateOperatorFunc)(unsigned long, unsigned long);}
\code{PredicateOperatorFunc fp = \&EqualInt4;}
\code{unsigned long column\_value;}
\code{unsigned long constant\_operand;}
\code{int result = fp(column\_value, constant\_operand);}
\vspace{2mm}

This C program can be modeled as follows.

\vspace{2mm}
\code{FuncType\_Compare : TYPE = (INT, INT, MEMORY) -> [INT, MEMORY]}
\code{EqualInt4 : FuncType\_Compare}
\code{= LAMBDA(value1 : INT, value2 : INT, in\_mem : MEMORY) : }
\code{LET}
\code{\qquad x = value1 - value2}
\code{IN (x, in\_mem);}
\vspace{2mm}
\code{LessthanInt8 : FuncType\_Compare}
\code{= LAMBDA(value1 : INT, value2 : INT, in\_mem : MEMORY) :}
\code{LET}
\code{x = IF value1 < value2 THEN 0 ELSE 1 ENDIF}
\code{IN (x, in\_mem);}
\vspace{2mm}
\code{addr\_EqualInt4, addr\_LessthanInt8 : INT;}
\code{ASSERT DISTINCT (addr\_EqualInt4, addr\_LessthanInt8);}
\code{fp : INT = addr\_EqualInt4;}
\code{column\_value : INT;}
\code{constant\_operand : INT;}
\code{mem : MEMORY;}
\code{result : [INT, MEMORY] = IF fp = addr\_EqualInt4}
\code{\qquad \qquad \qquad \qquad \qquad THEN EqualInt4(column\_value, constant\_operand, mem)}
\code{\qquad \qquad \qquad \qquad \qquad ELSIF fp = addr\_LessthanInt8}
\code{\qquad \qquad \qquad \qquad \qquad THEN LessthanInt8(column\_value, constant\_operand, mem)}
\code{\qquad \qquad \qquad \qquad \qquad ELSE (0, mem)}
\code{\qquad \qquad \qquad \qquad \qquad ENDIF;}
\vspace{2mm}

\subsection{More complicated structures}

\subsubsection{Loop}\label{sec:loop2}
Loops are difficult to model because CVC4-language does not allow loop iterators to be updated on themselves. But we can approximate to loop behavior via induction.

One straightforward intuition is unwinding the loop for exactly the number of iterations being executed at runtime. However, there are two potential problems. First, the number of iterations of a loop may be very large, which makes it impractical to unwind all iterations. Second, the number of iterations of a loop may dependent on runtime data which is unknown at the static analysis time. Therefore, we have to think of another idea to model loops in CVC4-language.

For a C program with a single loop, if a variable remains the same value before the loop, at the bottom of the loop body for all possible execution path, and after the loop, this variable can be considered as an invariant. This conclusion can be proved by induction. Therefore, each loop in the C program can be modeled with only one iteration. Note that we have to model all possible execution paths in the loop body in this one iteration to guarantee correctness. This modeling guarantees that all variables being labeled as invariants are indeed unchanged. 

In some cases, however, the modeling method presented above fails to recognize some invariants because the information presented by a single iteration is not enough. One example of these exceptions is listed below. In this example, \code{sum} has the same value at the second line and after the loop. But modeling with only one iteration cannot recognize this fact.

\vspace{2mm}
\code{int arr[5] = \{1, 2, 3, 4, 5\};}
\code{int sum = arr[0] + arr[1] + arr[2] + arr[3] + arr[4];}
\vspace{2mm}
\code{int i;}
\code{sum = 0;}
\code{for (i = 0; i < 5; ++i)}
\code{\{}
\code{\qquad sum += arr[i];}
\code{\}}
\vspace{2mm}

\subsubsection{Recursive function}\label{sec:recursion}
Because CVC4 doesn't support recursion, a recursive function has to be transformed into iteration. When it comes to loop, the problems are the same with those we have just described above.

\subsubsection{Address of members in structures}\label{sec:member addr}
In our method, we model the memory with an array for each C basic type, and associate an instance with an address by \texttt{ASSERT} statement. So each instance has only one address, including structures like tuple, record, and array. The reference to a member in an instance of these structures is a potential problem.

One possibility might be whenever the adress of a structure's member is used, we simply give it an undefined value. However, if this address is dereferenced by an update operation later, CVC4 considers that all instance in the same memory array may be changed because this dereferenced address has a non-deterministic value. A naive fix-up is to assert that this address is distinct from all other addresses used in the same program. It, however, doesn't work in the following case.

\vspace{2mm}
\code{struct T1}
\code{\{}
\code{\qquad int idx;}
\code{\};}
\vspace{2mm}
\code{struct T2}
\code{\{}
\code{\qquad T1 t;}
\code{\};}
\vspace{2mm}
\code{int arr[5] = \{1, 2, 3, 4, 5\};}
\code{T1 t\_1;}
\code{t\_1.idx = 0;}
\code{T2 t\_2;}
\code{t\_2.t = t\_1;}
\vspace{2mm}
\code{T1 *tp = \&t\_2.t;}
\code{arr[tp->idx] = 0;}
\vspace{2mm}

In this program, \code{tp} actually points to \code{t\_1}. But since we cannot model \code{\&t\_2.t}, we assign an unknown value which is distinct from all other addresses to \code{tp}. Because \code{tp} is non-deterministic, so is \code{tp->idx}. As the result, all element in \code{arr} are considered to be possibly changed by the last statement even though only the first element should be changed according to the semantics of the original C program.

Our first attempt is to extend the address from an integer to a structure including a base address and an offset. Also, all the structs in C are modeled with tuples in CVC4-language such that the member of their instance can be accessed with an offset rather than a name.

\vspace{2mm}
\code{ADDRESS : TYPE = [\# base : INT, offset : INT \#];}
\code{Memory\_someStruct : TYPE = ARRAY INT OF SomeStructure;}
\code{addr : ADDRESS;}
\code{ASSERT(mem.Memory\_someStruct[addr.base] = someInstance);}
\vspace{2mm}

In the code above, \code{base} is the index of the target instance in memory array and \code{offset} indicates the number of the target member in this instance. Unfortunately, we do not use this method because
\code{mem.Memory\_someStruct[addr.base].(addr.offset)} is an illegal expression in CVC language. Another problem is that if a function receives a structure member address and a new value for an update, unless the type of the host structure is known in the context, this function doesn't know which memory array to update. A simple C example is given below:

\vspace{2mm}
\code{void update(int *ip, int new)}
\code{\{}
\code{\qquad *ip = new;}
\code{\}}
\code{update(someInstance.int\_member, 0);}
\vspace{2mm}

Focusing on these two issues, our second attempt is to store structure members separately. The main idea here is to use a memory array for each basic types and put the structure members in these arrays corresponding to their types.

For each record \code{R : TYPE = [\# t1 : T1, t2 : T2, ..., tn : Tn \#]} where \code{Ti}$(0<i\le n)$ is a basic type, the address structure is \code{ADDRESS : TYPE = [\# base : INT, offset : INT \#]}, just like our first attempt. The index of memory arrays, however, is not an integer but an instance of type \code{ADDRESS}, which means that for an instance \code{r} of \code{R} whose base address is \code{k}, the value of \code{ti} in \code{r} resides at \code{(k, i)} in the memory array for \code{Ti}.

\vspace{2mm}
\code{ADDRESS : TYPE = [\# base : INT, offset : INT \#];}
\code{R : TYPE = [\# t1 : T1, t2 : T2, ..., tn : Tn \#];}
\code{Memory\_R : TYPE = ARRAY INT OF R;}
\code{Memory\_T1 : TYPE = ARRAY ADDRESS OF T1;}
\code{ASSERT(mem.Memory\_R[k] = r)}
\code{addr : ADDRESS = [\# k, i \#];}
\code{ASSERT(mem.Memory\_T1[addr] = valur\_of\_ti);}
\vspace{2mm}

With this modeling, the target array of an update is known once we know the type of the structure member to be updated. Since the address structure is used as the index of memory arrays, there is no syntax problem. Unfortunately, two other problems emerge for more complex programs. First, we need a more complex address structure to model possibly nested C program structs. Second, storing structure members separately according to their types makes it difficult to compare two arrays of structures in their entirety, which is sometimes useful.

\section{Heuristic of modeling with weaker precision}\label{sec: heuristic}
In Section~\ref{sec:member addr}, we have seen that it is difficult to model address of members in structures, but it is possible that some program invariants are not influenced by this kind of address. In this situation, we would like partially model the program to find invariants not influenced by structures that we cannot model accurately. A simple example to illustrate our idea is given below:

\vspace{2mm}
\code{void foo(struct ComplexType c1, int i) \{}
\code{\qquad int *mp = \&c1.member1;}
\code{\qquad int cnt = 0;}
\code{\qquad while (cnt < 100) \{}
\code{\qquad \qquad if (i) \{}
\code{\qquad \qquad \qquad cnt++;}
\code{\qquad \qquad \qquad *mp += cnt;}
\code{\qquad \qquad \}}
\code{\qquad \}}
\code{\}}
\vspace{2mm}

In this example, we cannot model \texttt{mp} accurately, but \texttt{i} is not influenced by \texttt{mp} and the value of \texttt{i} is an invariant. If we can identify the part of the program not influenced by \texttt{mp}, we could model this part of the program and identify the invariant of \texttt{i}.

In general, we would like to model parts of a program and detect invariant properties as long as they are not influenced by complex structures that we cannot accurately model. Since there is always a complicated interrelationship among variables in programs, it is necessary to determine how variables are influenced by each other.

The rest of this section propose an initial idea of determining the influence among variables in a program and modeling the parts of the program not influenced by variables that we cannot model accurately.

\subsection{Methodology}
Before presenting the method to compute the interrelationship among variables, we first make some claims.

\vspace{2mm}
$Definition 1$: We define that a variable $v_i$ $depends$ on a variable $v_j$ if and only if the change to the value of $v_j$ may cause the change to the value of $v_i$. And we say that $v_i$ is polluted by an item $t$ if we cannot model $v_i$ accurately without knowing how to model $t$.

\vspace{2mm}
$Claim 1$: In any C program $P$, there is a variable $v$ being polluted by item $t$ iff $v$ $depends$ on $t$.

$Proof$:  First, we prove that if a variable $v$ is polluted by an item $t$, $v$ $depends$ on $t$ by contradiction. Suppose a variable $v$ is polluted by an item $t$ but does not $depend$ on $t$. $v$ doesn't $depend$ on $t$ means that the change to the value of $t$ will not influence the value of $v$. So $v$ either has a deterministic value or an unknown value, which means that we can model $v$ simply with this value. But we supposed that $v$ is polluted by $t$, which indicates that we cannot model $v$, so there is a contradiction. Therefore, if a variable $v$ is polluted by an item $t$, $v$ $depends$ on $t$

Now, we prove that if a variable $v$ $depends$ on an item $t$, $v$ is polluted by $t$ by contradiction. Suppose that there is a variable $v$ that depends on $t$ while $v$ is not polluted by $t$. $v$ depends on $t$ means that the change to the value of $t$ may cause the change to the value of $v$. So we cannot model $v$ accurately without knowing how to model $t$. But we've supposed that $v$ is not polluted, which means that we can model $v$, so there is a contradiction. Therefore, if some variable $v$ $depends$ on item $t$, $v$ is polluted by $t$.

\vspace{2mm}
$Claim 2$: In any C program $P$ without function call, there is a variable being polluted by an item $t$ only if $t$ appears in assignment statements.

$Proof$: The statements in program can be roughly classified into three categories, assignment statement, control statement, and variable definition. Note that we consider a compound statement as a series of separate statements here.

For $t$ appearing in variable definition, other variables will not be polluted because there is no other variable introduced in a variable definition.

For $t$ appearing in control statement as a condition, it influences the execution path rather than other variables. Suppose there is an if-else branching in $P$ as follows where $t$ appears in \code{condition 1}, because we cannot model $t$, we cannot model \code{condition 1} either. But for any variable in \code{statement $j$}, the only possible relationship between its value and $t$ is that its value, which can be modeled as either its original value or the value appears in \code{statement $j$}, may depend on the execution path. Since all the possible values of this variable can be modeled, it's not polluted by the item $t$.

\vspace{2mm}
\code{if (condition 1)}
\code{\{}
\code{\qquad statement 1;}
\code{\qquad statement 2;}
\code{\qquad ...}
\code{\qquad statement $k$;}
\code{\}}
\code{else}
\code{\{}
\code{\qquad statement $k$+1;}
\code{\qquad statement $k$+2;}
\code{\qquad ...}
\code{\}}
\vspace{2mm}

For $t$ appearing in assignment statement, however, other variables may be polluted because there is either an explicit or implicit dependency between them. We'll see which variables are polluted by $t$ if $t$ appears in assignment statements in the following claims.

\vspace{2mm}
$Observation$: Every single assignment in a C program without function call can be transformed into a series of simple assignment statements in the form of $x = op(y_1, y_2)$ or $*x = op(y_1, y_2)$ where $op$ is a single operator other than function call, and $x, y_1, y_2$ are simple variables. (The case of $x = op(y_1)$ and $*x = op(y_1)$ is even simpler and can be inferred from the case using binary operator directly.)

\vspace{2mm}
$Claim 3$: In simple assignment $x = op(y_1, y_2)$, $x$ may be polluted if either of $y_1$ and $y_2$ is polluted.

$Proof$: We prove this claim by contradiction. Suppose that in this assignment, $x$ is polluted while none of $y_1$ and $y_2$ is polluted. Since $x$ depends on $y_1$ and $y_2$, and both $y_1$ and $y_2$ are not polluted indicates that we can model them accurately,  we can model $x$ accurately with a given $op$. So $x$ is not polluted. But we supposed that $x$ is polluted, which is a contradiction. Therefore, $x$ may be polluted if either of $y_1$ and $y_2$ is polluted.

\vspace{2mm}
$Definition 2$: We define a variable $v$ to be the base variable pointed by a pointer $x$ iff the object pointed by $x$ is $v$ itself, or the pointed object is a member of $v$ and $v$ is not a member of any other variables.

\vspace{2mm}
$Claim 4$: In simple assignment $*x = op(y_1, y_2)$, the base variable pointed by $x$ may be polluted if either of $x, y_1, y_2$ is polluted.

$Proof$: We prove this claim by contradiction. The semantic of $*x = op(y_1, y_2)$ is assigning the value of $op(y_1, y_2)$ to the object pointed by $x$. Suppose that in this assignment, the base variable pointed by $x$ is polluted while none of $x$, $y_1$ and $y_2$ is polluted. All of $x$, $y_1$ and $y_2$ are not polluted indicates that we can model them accurately. Since the new value of the base variable depends on $y_1$ and $y_2$, now we know which object should be assigned with what value, which means that we can do this assignment deterministically. So the base variable pointed by $x$ is not polluted. But we supposed that the base variable is polluted, which is a contradiction. Therefore, the base variable pointed by $x$ may be polluted if either of $x, y_1$ and $y_2$ is polluted.

\subsubsection{Constructing interrelationship graph}
Now we present a way to decide which variables in a C program are polluted by a given item $t$. To find all the variables being polluted by $t$ in a program $P$, we construct a directed graph $G=(V, E)$ in the following way.

\begin{enumerate}
\item{Inline} all the functions used in $P$.

\item{Transform} all the assignments to a series of simple assignment statements. Let $P'$ be the new program.

\item{For} each variable $v_i$ in $P'$, add a vertex $u_i$ to $G$.

\item{For} each simple assignment $x = op(y_1, y_2)$, add edges $(y_1, x), (y_2, x)$ to $G$.

\item{For} each simple assignment $*x = op(y_1, y_2)$ where $v$ is a base variable pointed by $x$, add edges $(y_1, v), (y_2, v), (x, v)$ to $G$.
\end{enumerate}

\vspace{2mm}
With this directed graph $G$ representing the interrelationship among variables in $P$, we can label all the variables that are polluted by a given item $t$ that we cannot model as follows.
\begin{enumerate}
\item{For} each assignment statement in $P$, if $t$ appears on the right-hand side, label the vertex corresponding to the variable on the left-hand side if it is a single variable, or label the base variable pointed by it if it is a dereference.

\item{For} each assignment statement in $P$, if $t$ appears on the left-hand side as a pointer dereference, label the vertex corresponding to the base variable pointed by $t$.

\item{Run} $Breadth First Searching$ on $G$ from the vertices being labeled and label all the vertices it can reach.
\end{enumerate}

The edges in $G$ represents the dependencies among variables, and if a variable $v$ depends on an item $t$ that we cannot model, we cannot model $v$ accurately either. So $v$ is polluted by the item $t$. Therefore, when modeling the program $P$, 
we first assign an unknown value to all the variables being labeled. Second, we remove all the statements involving dereference of unknown variables. Finally, we replace all the control condition, such as condition of if statement, involving an unknown variable with an unknown value of boolean type.

\subsubsection{A problem in this method}
\vspace{2mm}
There is, however, a problem with function inlining. Considering function inlining, functions can be classified into three categories: regular functions, recursive functions, and library functions. Regular functions can be inlined in the regular way, while it is problematic to inline recursive functions and library functions. The problem of recursive function inlining is that there is still a function call after inlining. And library functions cannot be easily inlined because we do not have the source code of them.

Our first attempt is to inline the recursive function but set all the branch conditions in it to be non-deterministic, and ignore the recursive call. The reason is that any variable actually changed by a function call should be assigned a new value in some statement other than the recursive call in the function body. But a counterexample is listed below.

\vspace{2mm}
\code{void RecursiveFunction(int *intArray, int index)}
\code{\{}
\code{\qquad intArray[index] = 0;}
\code{\qquad if (index >= 2)}
\code{\qquad \{}
\code{\qquad \qquad RecursiveFnction(intArray, index - 2);}
\code{\qquad \}}
\code{\}}
\vspace{2mm}

Obvious that by inlining only one call, we cannot reach all the indices where the elements of array are changed. Someone may think that we could set the arguments being modified in the recursive call to be non-deterministic, but it is possible that some of those arguments are pointers and a non-deterministic pointer could cause more pollution.

The problem to library functions is the same because we do not know the semantics of these functions. Given the semantics of them, maybe we could do something further.

\subsection{Formal proof}
In this section, we give a proof that the variables whose corresponding vertices are not labeled in $G$ will not be polluted by a given item. To prove this conclusion, we first give some claims.

\vspace{2mm}
$Claim 5$: A variable $v_i$ $depends$ on a variable $v_j$ if and only if there is a path in $G$ from $u_j$ to $u_i$.

$Proof$: First, we prove that if there is a path in $G$ of length $k$ from $u_j$ to $u_i$, $v_i$ $depends$ on $v_j$. We prove this conclusion by induction of $k$.

Base case: When $k = 1$, $u_j$ is adjacent to $u_i$, which means that there is an edge $(u_j, u_i)$ in $G$. Recall the construction of $G$, in either case where we add an edge $(u_j, u_i)$ to $G$, the value of $v_i$ depends on the value of $v_j$.

Induction hypothesis: Suppose a path of length $k$ from $u_j$ to $u_i$ in $G$ implies that $v_i$ $depends$ on $v_j$.

Induction step: Let $p = u_{n_1}, u_{n_2}, u_{n_3}, ..., u_{n_k}, u_{n_{k+1}}$ be a path of length $k$+1 from $u_{n_1}$ to $u_{n_{k+1}}$. Since $p$ is a path of length $k$+1, $p' = u_{n_1}, u_{n_2}, u_{n_3}, ..., u_{n_k}$ is a path of length $k$. Because we supposed that a path of length $k$ from $u_j$ to $u_i$ in $G$ implies that $v_i$ $depends$ on $v_j$, and an edge $(u_{n_k}, u_{n_{k+1}})$ indicates that the value of $v_{n_{k+1}}$ depends on the value of $v_{n_k}$ as we've shown in the base case, we know $v_{n_k}$ depends on $v_{n_1}$ and $v_{n_{k+1}}$ depends on $v_{n_k}$. So $v_{n_{k+1}}$ depends on $v_{n_1}$.

Now we prove that if $v_i$ $depends$ on some variable $v_j$, there is a path in $G$ from $u_j$ to $u_i$ by contradiction.

Suppose that there is variable $v_i$ depends on some variable $v_j$ while there is no path from $u_j$ to $u_i$ in $G$, we'll show a contradiction. Because $v_i$ depends on $v_j$, $v_i$ depends directly on either $v_j$ itself or another variable $v_m$ depending on $v_j$. If $v_i$ depends directly on $v_j$, there must be an edge from $u_j$ to $u_i$ according to our construction. So there is a path of length 1 from $u_j$ to $u_i$. But we supposed that there is no path from $u_j$ to $u_i$, which is a contradiction. If $v_i$ depends directly on another variable $v_m$ that depends on $v_j$, there must be no path from $u_j$ to $u_m$ because otherwise there is an edge from $u_m$ to $u_i$ according to our construction and thus a path from $u_j$ to $u_i$ through $u_m$. Likewise, $v_m$ depends directly on either $v_j$ itself or another variable depending on $v_j$. If we trace back like this, we can finally find a vertex $v_d$ which depends directly on $v_j$ yet there is no edge from $u_j$ to $u_d$. But according to our construction, there must be an edge $(u_j, u_d)$, which is a contradiction. Therefore, if $v_i$ $depends$ on some variable $v_j$, there is a path in $G$ from $u_j$ to $u_i$.

\vspace{2mm}
$Claim 6$: A variable $v_i$ in $P$ is polluted by $t$ if and only if there is a path in $G$ from a labeled vertex to the vertex $u_i$.

Before we prove this claim, we first show that the initial labeled vertices are polluted. According to our construction of $G$, the initial labeled vertices are those corresponding to variables which appear on the left-hand side of some assignment statements where $t$ appears on the right-hand side. These variables are polluted because their values depend on the value of $t$, and we cannot model them without knowing how to model $t$.

$Proof$: First, we prove that if there is a path in $G$ from the vertex $u_i$ to a labeled vertex $u_p$, $v_i$ in $P$ is polluted. Because there is a path in $G$ from the vertex $u_i$ to a labeled vertex $u_p$, the value of $v_i$ depends on the value of $v_p$. Since $v_p$ is polluted by $t$ as we showed above, which means that $v_p$ depends on $t$, $v_i$ also depends on $t$. So $v_i$ is polluted.

Now, we prove that if a variable $v_i$ in $P$ is polluted by $t$, there is a path in $G$ from a labeled vertex $u_p$ to $u_i$.

Because $v_i$ is polluted by $t$, $v_i$ depends on $t$. So $v_i$ depends directly on either $t$ itself or another variable $v_m$ depending on $t$. If $v_i$ depends directly on $t$, $u_i$ is an initial labeled vertex according to our construction. So there is a path of length 0 from a labeled vertex to $u_i$. If $v_i$ depends on another variable $v_m$ depending on $t$, there is a path from $u_m$ to $u_i$. Likewise, $v_m$ depends directly on either $t$ itself or another variable depending on $t$. If we trace back like this, we can finally find a vertex $v_d$ which depends directly on $t$. According to our construction, $v_d$ is an initial labeled vertex, so there is a path from a labeled vertex to $v_i$.

\vspace{2mm}
So far, we have shown that a vertex $u_i$ is accessible from a labeled vertex iff the variable $v_i$ is polluted by the item $t$. Because we cannot model these variables, we label all the variables accessible from a labeled vertex and assign a non-deterministic value to them. Also, considering that some of these variables may be pointers and it's unsafe to dereference a non-deterministic pointer, we remove the assignment statements where a dereference of an unknown pointer appears. 
It is safe to remove a dereference of non-deterministic pointer because the base variable pointed by it is labeled and assigned a non-deterministic value. Finally, the non-deterministic value of labeled variables will not influence unlabeled variables because all the vertices accessible from them have already been labeled, which means that the value of unlabeled variables do not depend on them. Therefore, it is safe to make them non-deterministic.

\subsection{Example}
As an example, we present a function extracted and simplified from one of our tiny prototype. The item that we cannot model is \code{\&((*predicates)[i])} which appears in \code{SequentialScan}.

\vspace{2mm}
\code{int SequentialScan(int scan\_direction,}
\code{\qquad \qquad \qquad \qquad int num\_predicates,}
\code{\qquad \qquad \qquad \qquad Predicate** predicates,}
\code{\qquad \qquad \qquad \qquad const TableHeader* schema)}
\code{\{}
\code{\qquad column\_offset = 0;}
\code{\qquad for (i = 0; i < schema->num\_columns; ++i)}
\code{\qquad \{}
\code{\qquad \qquad column\_type = schema->column\_definitions[i].column\_type;}
\code{\qquad \qquad if (column\_type == DATATYPE\_INT4)}
\code{\qquad \qquad \{}
\code{\qquad \qquad \qquad row\_values[i] = *(int*)\&(row\_data[column\_offset]));}
\code{\qquad \qquad \qquad column\_offset += 4;}
\code{\qquad \qquad \}}
\code{\qquad \qquad else if (column\_type == DATATYPE\_INT8)}
\code{\qquad \qquad \{}
\code{\qquad \qquad \qquad row\_values[i] = *(long*)(\&(row\_data[column\_offset]));}
\code{\qquad \qquad \qquad column\_offset += 8;}
\code{\qquad \qquad \}}
\code{\qquad \}}
\code{\qquad column\_value = 0;}
\code{\qquad current\_predicate = NULL;}
\code{\qquad for (i = 0; i < num\_predicates; ++i)}
\code{\qquad \{}
\code{\qquad \qquad current\_predicate = \&((*predicates)[i]);}
\code{\qquad \qquad column\_value = row\_values[current\_predicate->column\_id];}
\code{\qquad \}}
\code{\qquad return 0;}
\code{\}}
\vspace{2mm}

Given this program, we first transform each assignment to a series of simple assignment statements as follows. And then construct a graph according to the variables and dependencies among them.

\vspace{2mm}
\code{int SequentialScan(int scan\_direction,}
\code{\qquad \qquad \qquad \qquad int num\_predicates,}
\code{\qquad \qquad \qquad \qquad Predicate** predicates,}
\code{\qquad \qquad \qquad \qquad const TableHeader* schema)}
\code{\{}
\code{\qquad column\_offset = 0;}
\code{\qquad for (i = 0; i < schema->num\_columns; ++i)}
\code{\qquad \{}
\code{\qquad \qquad TableHeader schema\_inst = *schema;}
\code{\qquad \qquad ColumnDefinition *col\_def;}
\code{\qquad \qquad col\_def = schema\_inst.column\_definitions;}
\code{\qquad \qquad ColumnDefinition col\_def\_i;}
\code{\qquad \qquad col\_def\_i = col\_def[i];}
\code{\qquad \qquad column\_type = col\_def\_i.column\_type;}
\code{\qquad \qquad if (column\_type == DATATYPE\_INT4)}
\code{\qquad \qquad \{}
\code{\qquad \qquad \qquad int *row\_value\_i;}
\code{\qquad \qquad \qquad row\_value\_i = row\_values + i * sizeof(int);}
\code{\qquad \qquad \qquad *row\_value\_i = row\_data[column\_offset];}
\code{\qquad \qquad \qquad column\_offset += 4;}
\code{\qquad \qquad \}}
\code{\qquad \qquad else if (column\_type == DATATYPE\_INT8)}
\code{\qquad \qquad \{}
\code{\qquad \qquad \qquad int *row\_value\_i = row\_values + i * sizeof(long);}
\code{\qquad \qquad \qquad *row\_value\_i = row\_data[column\_offset];}
\code{\qquad \qquad \qquad column\_offset += 8;}
\code{\qquad \qquad \}}
\code{\qquad \}}
\code{\qquad column\_value = 0;}
\code{\qquad current\_predicate = NULL;}
\code{\qquad for (i = 0; i < num\_predicates; ++i)}
\code{\qquad \{}
\code{\qquad \qquad Predicate *predicates\_arr;}
\code{\qquad \qquad predicates\_arr = *predicates;}
\code{\qquad \qquad Predicate predicate\_i;}
\code{\qquad \qquad predicate\_i = predicates\_arr[i];}
\code{\qquad \qquad current\_predicate = \&predicate\_i;}
\code{\qquad \qquad Predicate curr\_pred\_inst;}
\code{\qquad \qquad curr\_pred\_inst = *current\_predicate;}
\code{\qquad \qquad int col\_id;}
\code{\qquad \qquad col\_id = curr\_pred\_inst.column\_id;}
\code{\qquad \qquad column\_value = row\_values[col\_id];}
\code{\qquad \}}
\code{\qquad return 0;}
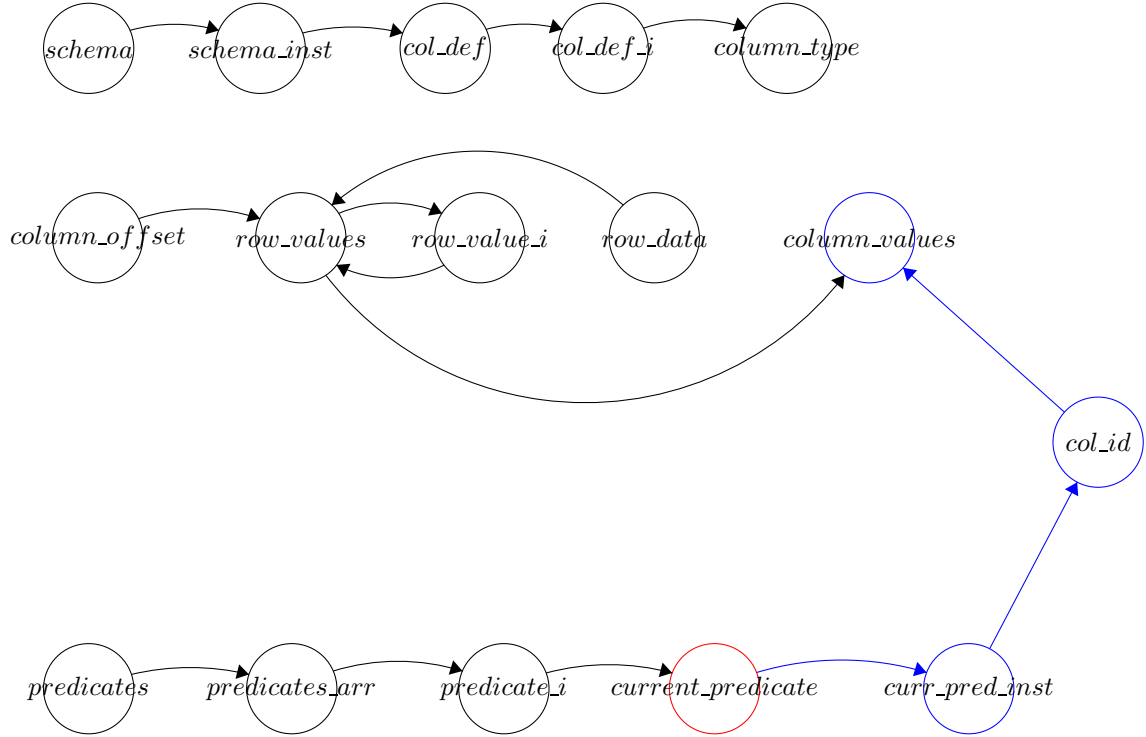
\code{\}}
\vspace{2mm}

\begin{center}
\begin{tikzpicture}[scale=0.2]
\tikzstyle{every node}+=[inner sep=0pt]
\draw [black] (6,-3.8) circle (3);
\draw (6,-3.8) node {$schema$};
\draw [black] (17.4,-3.8) circle (3);
\draw (17.4,-3.8) node {$schema\_inst$};
\draw [black] (29.7,-3.8) circle (3);
\draw (29.7,-3.8) node {$col\_def$};
\draw [black] (40.2,-3.8) circle (3);
\draw (40.2,-3.8) node {$col\_def\_i$};
\draw [black] (52.4,-3.8) circle (3);
\draw (52.4,-3.8) node {$column\_type$};
\draw [black] (6.6,-16.4) circle (3);
\draw (6.6,-16.4) node {$column\_offset$};
\draw [black] (20.1,-16.4) circle (3);
\draw (20.1,-16.4) node {$row\_values$};
\draw [black] (32,-16.4) circle (3);
\draw (32,-16.4) node {$row\_value\_i$};
\draw [black] (43.6,-16.4) circle (3);
\draw (43.6,-16.4) node {$row\_data$};
\draw [blue] (57.9,-16.4) circle (3);
\draw (57.9,-16.4) node {$column\_values$};
\draw [black] (6,-46.4) circle (3);
\draw (6,-46.4) node {$predicates$};
\draw [black] (19.5,-46.4) circle (3);
\draw (19.5,-46.4) node {$predicates\_arr$};
\draw [black] (33.6,-46.4) circle (3);
\draw (33.6,-46.4) node {$predicate\_i$};
\draw [red] (47.6,-46.4) circle (3);
\draw (47.6,-46.4) node {$current\_predicate$};
\draw [blue] (64.5,-46.4) circle (3);
\draw (64.5,-46.4) node {$curr\_pred\_inst$};
\draw [blue] (73.1,-30) circle (3);
\draw (73.1,-30) node {$col\_id$};
\draw [black] (8.742,-2.606) arc (105.68481:74.31519:10.942);
\fill [black] (14.66,-2.61) -- (14.02,-1.91) -- (13.75,-2.87);
\draw [black] (9.311,-15.131) arc (108.37316:71.62684:12.814);
\fill [black] (17.39,-15.13) -- (16.79,-14.4) -- (16.47,-15.35);
\draw [black] (22.611,-14.785) arc (112.98965:67.01035:8.805);
\fill [black] (29.49,-14.78) -- (28.95,-14.01) -- (28.56,-14.93);
\draw [black] (29.633,-18.215) arc (-63.29975:-116.70025:7.975);
\fill [black] (22.47,-18.21) -- (22.96,-19.02) -- (23.41,-18.13);
\draw [black] (22.163,-14.229) arc (130.68258:49.31742:14.861);
\fill [black] (22.16,-14.23) -- (23.1,-14.09) -- (22.44,-13.33);
\draw [black] (56.238,-18.894) arc (-37.62879:-142.37121:21.765);
\fill [black] (56.24,-18.89) -- (55.35,-19.22) -- (56.15,-19.83);
\draw [black] (8.841,-45.45) arc (103.3938:76.6062:16.873);
\fill [black] (16.66,-45.45) -- (16,-44.78) -- (15.76,-45.75);
\draw [black] (22.254,-45.225) arc (107.19535:72.80465:14.53);
\fill [black] (30.85,-45.22) -- (30.23,-44.51) -- (29.93,-45.47);
\draw [black] (36.407,-45.355) arc (105.08502:74.91498:16.11);
\fill [black] (44.79,-45.35) -- (44.15,-44.66) -- (43.89,-45.63);
\draw [blue] (50.397,-45.322) arc (106.70455:73.29545:19.668);
\fill [blue] (61.7,-45.32) -- (61.08,-44.61) -- (60.79,-45.57);
\draw [blue] (65.89,-43.74) -- (71.71,-32.66);
\fill [blue] (71.71,-32.66) -- (70.89,-33.13) -- (71.78,-33.6);
\draw [blue] (70.86,-28) -- (60.14,-18.4);
\fill [blue] (60.14,-18.4) -- (60.4,-19.31) -- (61.07,-18.56);
\draw [black] (20.216,-2.782) arc (103.75013:76.24987:14.027);
\fill [black] (26.88,-2.78) -- (26.23,-2.11) -- (25.99,-3.08);
\draw [black] (32.409,-2.541) arc (105.75205:74.24795:9.361);
\fill [black] (37.49,-2.54) -- (36.86,-1.84) -- (36.59,-2.8);
\draw [black] (42.874,-2.461) arc (108.60453:71.39547:10.74);
\fill [black] (49.73,-2.46) -- (49.13,-1.73) -- (48.81,-2.68);
\end{tikzpicture}
\end{center}

According to our construction, the initial labeled vertex is \code{current\_predicate} because \linebreak \code{\&((*predicates)[i])} appears in the statement \code{current\_predicate = \&((*predicates)[i]);}. We run $Breadth First Search$ on $G$ from the vertex $current\_predicate$ and label all the vertices being visited, which is $curr\_pred\_inst$, $col\_id$ and $column\_value$. we replace them with non-deterministic values.

One potential problem of this method is that the transformation from each assignment to a series of simple assignment statements introduces more variables, which increases the scale of the graph. Also, when it is applied to larger and more complex programs, function inlining may be unpractical.

\section{Conclusion}
This paper describes our attempt of detecting program invariants with sophisticated analysis by program verification tools. This work is motivated by the fact that some program invariants that enable significant performance improvement can be identified by domain experts of the programs but not compilers. One reason to this limitation of compilers is that compilers usually do not perform expensive analysis due to performance concerns. 

We described four tools we have tried for invariant detection: one SMT solver and three C program analysis tools based on SMT or SAT solvers. We noticed that these C program tools sometimes take an approximation to the examined property and only tools taking over-approximation is safe to use for invariant detection. These C program tools either take under-approximation or are not robust enough for large programs.

On the other hand, it is hard for SMT solver to model some complex C structures. For one of these tools---CVC4, we described how we can model program behavior in its input language and discussed challenges of modeling loops and addresses of structure members. Finally, we gave an initial idea for identify a subset of invariants using CVC4 when some C structures in the program cannot be modeling accurately.

\bibliographystyle{plain}
\bibliography{review}

\end{document}